\documentclass[final,5p,times,twocolumn]{elsarticle}

 \usepackage{graphics}
 \usepackage{graphicx}
 
\usepackage{amssymb}
\usepackage{amsthm}

\usepackage{algorithm}
\usepackage{algorithmicx}
\usepackage{algpseudocode}

\makeatletter
\def\ps@pprintTitle{%
  \let\@oddhead\@empty
  \let\@evenhead\@empty
  \def\@oddfoot{\reset@font\hfil\thepage\hfil}
  \let\@evenfoot\@oddfoot
}
\makeatother

\journal{}

\begin{document}

\makeatletter
\newlength \figwidth
\if@twocolumn
  \setlength \figwidth {1\columnwidth}
\else
  \setlength \figwidth {0.8\textwidth}
\fi
\makeatother

\begin{frontmatter}




\title{Tag-based Semantic Website Recommendation for Turkish Language}


\author{Onur Y{\i}lmaz }
\address{Computer Engineering Department, Middle East Technical University, Ankara, Turkey}
\ead{<onur@onuryilmaz.me>}


\begin{abstract}

With the dramatic increase in the number of websites on the internet, tagging has become popular for finding related, personal and important documents. When the potentially increasing internet markets are analyzed, Turkey, in which most of the people use Turkish language on the internet, found to be exponentially increasing. In this paper, a tag-based website recommendation method is presented, where similarity measures are combined with semantic relationships of tags. In order to evaluate the system, an experiment with 25 people from Turkey is undertaken and participants are firstly asked to provide websites and tags in Turkish and then they are asked to evaluate recommended websites.

\end{abstract}


\end{frontmatter}


\section{Introduction}
\label{}

Tags are defined as non-hierarchical keyword or term assigned to a piece of information. Using tags helps to describe a document and allows it to be found again by browsing or searching. Reasons behind tagging can be various such as categorizing, memorizing, archiving and sharing. With the rise of Web 2.0, users start to tag items for not themselves but also for sharing with others. This created the area of collaborative tagging, known as folksonomy, in which users collectively classify and find information. 

With the dramatic increase in the number of the websites on the internet (7.14 billion pages as of December 2012\footnotemark), \footnotetext[1]{worldwidewebsize.com} finding the important and related information has become difficult. This difficulty created need for social bookmarking networks where collaborative tagging of web resources are made. In social bookmarking environments, recommendation systems are implemented to suggest new and focused resources to users. 

Users in social bookmarking networks tag the resources themselves and use their own language in general. When the potentially increasing internet markets are analyzed, it is found out that Turkey shows an exponential increase in the last years \cite{ispa2010}. In spite of this increase, English Proficiency Index lists Turkey as 32nd country with the mark of low proficiency \cite{efepi2012}. This figure shows that most of the internet users in Turkey cannot use English efficiently. In addition, it can be inferred that internet users in Turkey, tag resources using Turkish which is a very different language than English. Difference between these languages mostly based on agglutinative property and grammatical structure \cite{frankfurt2001}. Thus, there is necessity of recommendation systems that incorporate Turkish language and consider usage styles of Turkish people. 

Considering this necessity, in this paper a Turkish-language tag-based recommendation system which is based on similarity, tag weight, tag popularity; where semantic properties of tags are taken into account is proposed. In this method, tag weight is used for measuring tag representativeness; and tag popularity is used for checking frequency. Other than semantics, most of the calculation method is adapted from paper of Durao \& Dolog \cite{durao2012personalized}. Semantic properties of tags included by the help of a synonym dictionary for Turkish.  In addition, spell-checking and stemming operations are done on tags to create a semi-controlled environment. 

Goal of this paper is analyzing whether tags can be utilized for personal recommendations within Turkish language. In order to measure the performance of this method, an experiment with the participation of random internet users is undertaken.  With this experiment, it is shown that more than 70 \% of the recommended websites are accepted by users.    

Main contribution of this paper is combining well-known similarity measures and calculations with a Turkish semantics analysis in which directly meanings of words are used instead of extracting patterns or topics.

This paper is organized so that, in the second part, related work is presented and in the third part, problem definition and implemented algorithm is explained. In the fourth part experiment is discussed and in the next parts conclusion and future works are mentioned.

\section{Related Work}
\label{}
Considering the dramatic increase in the number of pages on the internet, finding the related information and websites has become more and more difficult. As Nakamoto et al. mentioned on their paper, collaborative filtering was found to be a solution for finding the related and personal information \cite{nakamoto2007tag}. In collaborative filtering, community and their inputs are used for finding and matching similar users; however, this method does not consider the context of the resources.

When the social communities over the internet are thought, it must be mentioned that there are also social tagging networks such as del.icio.us\footnotemark \footnotetext[2]{delicious.com}, AddThis\footnotemark \footnotetext[3]{addthis.com} or BlinkList\footnotemark \footnotetext[4]{blinklist.com}. As Cattuto et al. mention discovery of concept hierarchies can be applied on these social tagging systems and these hierarchies can be mapped to WordNet \cite{cattuto2008semantic}. Moreover, in ConTag extracting topics from tags is implemented to reveal semantic relationships \cite{adrian2007contag}. However, in this paper instead of patterns, topics or concept hierarchies, directly meaning of words are used for semantic analysis.

When the different similarity measurements are surveyed, Durao \& Dolog implemented a combination of basic similarity, tag popularity, tag representativeness and tag-user affinity \cite{durao2012personalized}. Implementing these measurements, they have reached 60 \% acceptance of recommended websites. In this paper an adapted version of their calculation method is implemented considering semantic properties of tags and a higher acceptance level is achieved.

Considering the current status of related work, in this paper, appropriate approaches of different models are combined and applied on Turkish language. Within the proposed model, tag popularity, representativeness, semantic relations and their similarity are taken into account to provide personal recommendations. In this model, similarity calculations and measures which are commonly accepted and applied are combined with a basic but beneficial semantic relationship implementation for Turkish language.

\section{Problem Definition and Algorithm}
\subsection{Task Definition}

In this paper, a Turkish language tag-based website recommendation system is implemented. In this system, inputs from internet users are collected as $<$website, tag$>$ pairs and new websites which is expected to be interesting, related and personal to the users are found. In other words, this model should recommend websites that the user wants to use in the future, or already using and finds them interesting. The difficulty in this recommendation system resulted from two main reasons. Firstly, there are billions of people with different interests, backgrounds; internet usage habits and expectations from recommended websites. Secondly, people tag resources for different purposes such as categorizing, memorizing, and archiving or for just being asked to do. Some examples of different purposes are provided in Table~\ref{table31}. This tagging activity is used as a mapping function applied on a sample set of websites. Therefore, achieving high level of user satisfaction in this area can be classified as a challenging task.

\begin{table}[ht]
\caption{Example purposes of tagging} 
\centering  
\includegraphics[width=\figwidth]{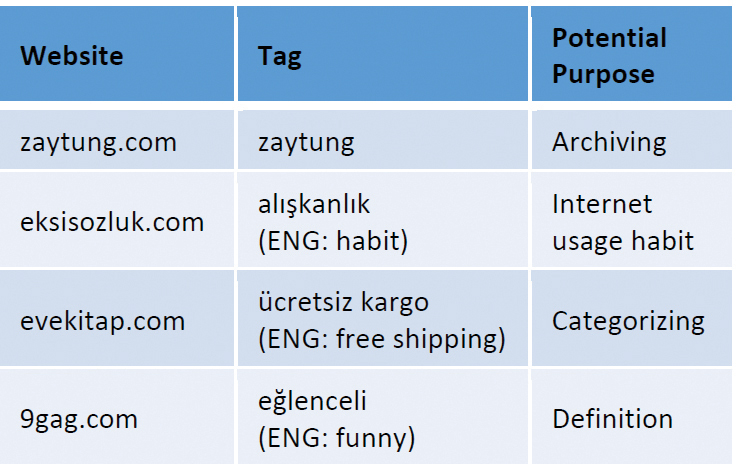}
\label{table31} 
\end{table}

 \subsection{Algorithm Description}
 \subsubsection{Preprocessing of Data}
 
In order to overcome differences caused by user inputs, firstly all tags are converted to lowercase letters without any information loss. Secondly, considering the potential of misspelling, spell-check is made on the tags. Since tags are short and independent keywords, one-edit distances of noisy channel approach is used: (1) add a single letter, (2) delete a single letter, (3) replace one letter and (4) transpose two letters \cite{brill2000improved}. In this spell checking, one edit distant words are created and checked whether they occur in a sample of Turkish National Corpus \cite{aksanconstruction}. If they occur in corpus, they are directly converted to the estimated corrected ones and used in the further steps.  

Thirdly, website URLs are processed considering the different representations of the same web resources. Since there are many different web technologies used on the internet, there could be different user inputs meaning the same homepage of the website. In addition, implementing HTTPS can yield different URLs. Therefore, all web page links provided by users are preprocessed as some examples provided in Table~\ref{table32}.

\begin{table}[ht]
\caption{Examples of URL correction} 
\centering  
\includegraphics[width=\figwidth]{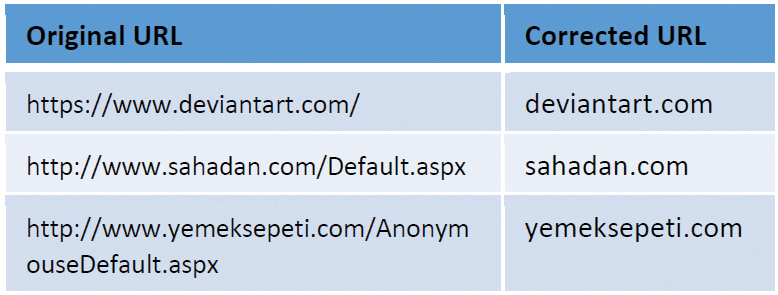}
\label{table32} 
\end{table}

Fourthly, considering Turkish language as an agglutinative language, effect of inflectional affixes tried to be minimized. There are many possibilities of a word to have affixes and change in meaning in Turkish; however, they have closely the same meaning in tag environment. These affixes are removed and stems of the words are used as tags so that relation between the words created from the same stem will be kept alive. Although stemming seems to cause information loss, as can be seen from the examples provided in Table~\ref{table33}, stemming yielded better tags for further steps.
 
 \begin{table}[ht]
\caption{Examples of tag stemming} 
\centering  
\includegraphics[width=\figwidth]{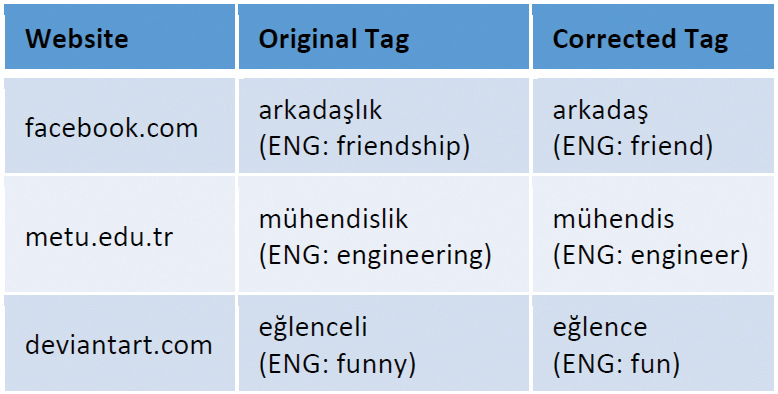}
\label{table33} 
\end{table}

 \subsubsection{Semantics Analysis}
 
 After preprocessing the data, all website names and tags converted so that they will reveal relationships between them more easily. In this step, semantic relations between tags are tried to be exploited. In order to understand and connect related tags, an open source Turkish thesaurus project\footnotemark \footnotetext[5]{github.com/maidis/mythes-tr} is used. From this project, all $<$word, synonym$>$ pairs are extracted and saved as SYNONYM-LIST. On the other hand all user inputs of $<$user, site, tag$>$ are held in ALL-DATA; and the algorithm provided in Algorithm~\ref{tagalgorithm} is applied. With this algorithm one-distance related words are found and they are added as they were inputs provided by users.

\begin{algorithm}
\caption{Algorithm for finding tag synonyms}
\label{tagalgorithm}
\begin{algorithmic}
\ForAll{$tag \in$ ALL-DATA} 
    \ForAll{$<tag, synonym>$ $\in$ SYNONYM-LIST} 
    \If {$ synonym \in$ ALL-DATA}
        \State Add $<user, site,synonym>$ to ALL-DATA 
    \EndIf
    \EndFor
\EndFor
\end{algorithmic}
\end{algorithm}
 
In order to present the effect of this algorithm, in experiment stage, this algorithm yielded 68 new $<$user, site, tag$>$ tuples whereas ALL-DATA have 366 tags. It means 18 \% of increase in dataset, in other words 18 \% more tags provided by users to define their websites. However, these added tags do not only define websites but also connects the related tags used by other participants and create an environment where all users provide tags and their potential meanings which other people may have already used. Relation between tags and added tuples are illustrated in Table~\ref{table34}.

\begin{table}[ht]
\caption{An example of semantics analysis algorithm} 
\centering  
\includegraphics[width=\figwidth]{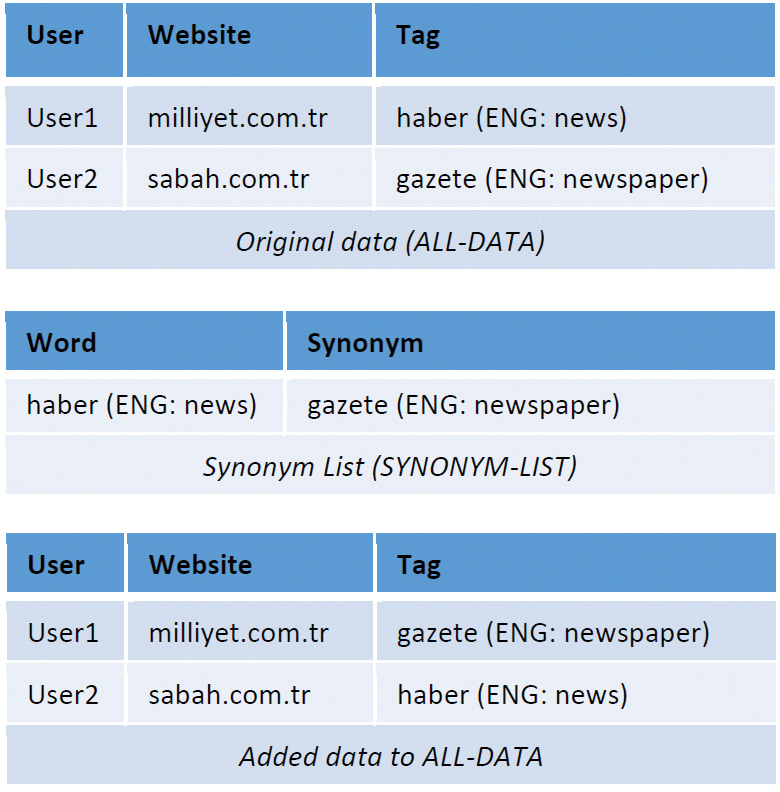}
\label{table34} 
\end{table}

 Although pseudo code seems short, this algorithm have time complexity of $O(|$ALL-DATA$|^2$ x $|$SYNONYM-LIST$|)$. Since the used thesaurus has only 125.022 $<$word, synonym$>$ pairs it did not create a time related problem; however, when this method is used with large thesauruses it can take more time to find all related tags and synonyms.

 \subsubsection{Similarity Calculations}
 
 Although related tags are found in semantic analysis, in order to provide personal recommendations to users, similarities between $<$user, website, tag$>$ pairs must be found. While calculating similarity between tuples, an adapted version of calculation presented in the paper of Durao \& Dolog is implemented \cite{durao2012personalized}. Considering the fact that some of the tags are added by semantic analysis; affinity of tags, in other words users’ tendency to use a tag, is removed from their calculation method and not implemented in this method. 

    Firstly, tag popularity is used for measuring how often the tag is used by users. Therefore, it is calculated by number of occurrences of the related tag divided by the total number of websites.
    
 \begin{figure}[h!]
 \centering
\includegraphics[width=\figwidth]{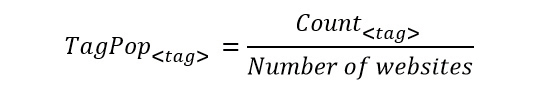}
 \end{figure}

Secondly, tag representativeness is implemented to measure how well the tags represent the documents. With this reasoning it is calculated by dividing $<$website, tag$>$ occurrences to the number of tag occurrences.

 \begin{figure}[h!]
 \centering
\includegraphics[width=\figwidth]{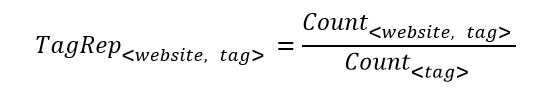}
 \end{figure}

Thirdly, cosine similarity between websites are implemented for each website where tags are used as vectors and cosine similarity between these vectors are calculated. In other words, cosine similarity sums the co-occurrences of tags in both websites in this method.

Using these measures, for each website, a rating is calculated as following:

 \begin{figure}[h!]
 \centering
\includegraphics[width=\figwidth]{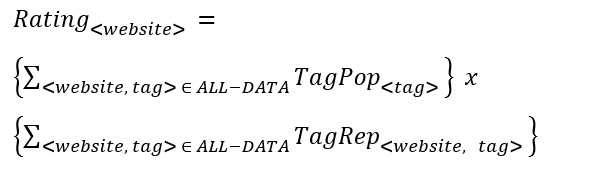}
 \end{figure}

Using these website ratings, similarity between them is calculated as:
 \begin{figure}[h!]
 \centering
\includegraphics[width=\figwidth]{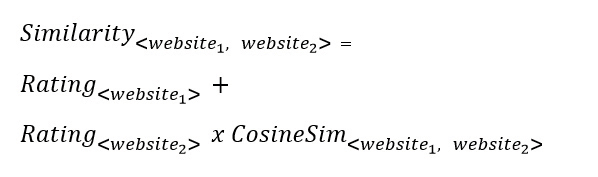}
 \end{figure}

 \subsubsection{Recommendation Extraction}
 
\begin{figure}[h!]
\centering
\includegraphics[width=\figwidth]{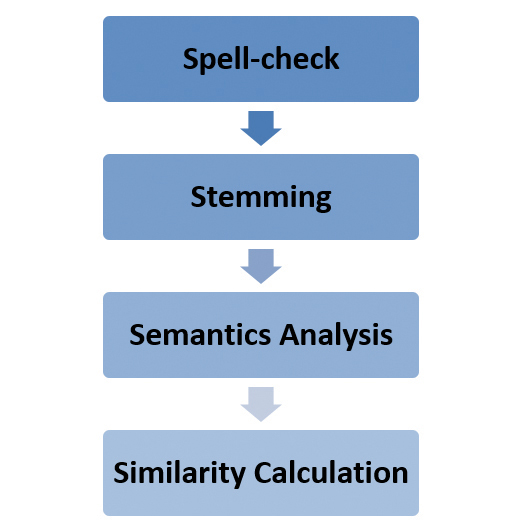}
\caption{Summary of approach}
\label{approach}
\end{figure}

 All steps of the proposed method can be summarized in Figure~\ref{approach}. After calculating similarity between all websites, the websites are sorted according to their calculated similarity value and they are grouped according to users. Finally, a list of recommended websites for each user is extracted and these lists are sorted from greater similarity to lower. Therefore, for any application, any number of recommended websites can be taken from the top of these lists.
 
\section{Experimental Evaluation}

\subsection{Methodology}

\begin{figure}[h!]
\centering
\includegraphics[width=\figwidth]{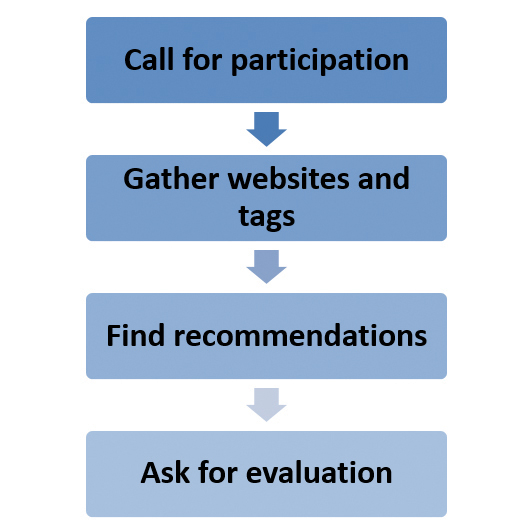}
\caption{Experiment steps}
\label{experiment}
\end{figure}

Evaluation of the model is based on how the users are satisfied with the recommended websites. In order to collect this feedback, an experiment is undertaken. In this experiment, firstly users are called for participation. To have a good sample of users with different backgrounds, a general purpose internet portal, namely Ek\c{s}i Duyuru\footnotemark \footnotetext[6]{eksiduyuru.com}, is used for call. These people are invited to a webpage\footnotemark \footnotetext[7]{bit.ly/oneri-sistemi} where they need to provide websites and tags in Turkish. In total, 25 users provide 122 websites and 366 tags in this stage. Following this action, algorithm steps mentioned in Section 3 are applied and recommended websites are extracted. Then for each participant, recommended websites are sent for evaluation. This evaluation is made by asking if these new websites can be accepted as interesting, useful or relevant to them. If the recommended website falls into one of these sets, participants are asked to choose as "Accepted" otherwise "Rejected" on an online poll\footnotemark \footnotetext[8]{bit.ly/oneri-degerlendirme}. Although 25 users participated to the prior steps, only 20 of them participated to the evaluation stage. Steps of the experiment can be diagrammed in Figure~\ref{experiment}.

\subsection{Expected Results}
Although a controlled environment is created by spell-checking and correcting URLs there are still human aspect in the proposed recommendation system. Considering the mentioned potential purposes of tagging in Section 3.1 and diversity of expectations from a recommendation system, it is not expected to have 100 \% acceptance level. However, it is a natural expectation to have at least half of the recommended websites as accepted \cite{durao2012personalized}. It is thought so because when the acceptance level is less than 50 \%, proposed model recommends not attractive or not related in most of the cases. 

In this experiment, user acceptance level is measured on two subsets of the recommended websites. Top five recommended documents are sent to users for evaluation in random order. Then their acceptance level in top five and top three are analyzed. As mentioned, both of these acceptance levels are expected to be over 50 \% and top three recommendations should achieve better.

\subsection{Results}
\begin{figure*}
  \includegraphics[width=\textwidth]{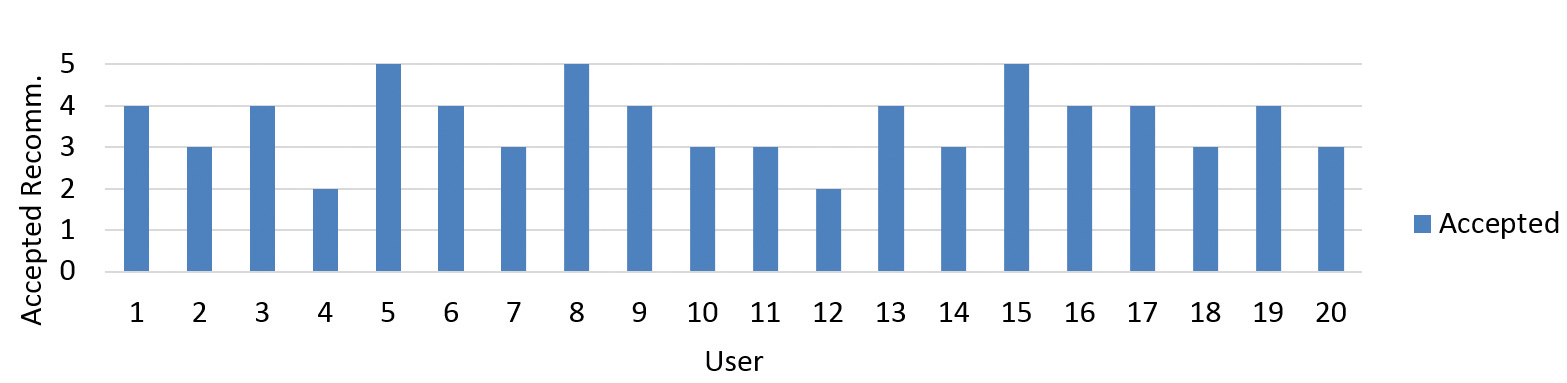}
  \caption{Accepted recommendations by each user (5 Recommendations)}
  \label{result-5-1}
\end{figure*}

Evaluation of the top 5 recommendations can be seen in Figure~\ref{result-5-1} for each user with the number of accepted websites. This figure shows at least 2 of the recommendations are accepted and 3 of the users have accepted all recommended websites.

    When the number of acceptance levels are further analyzed, 3.6 of 5 websites are accepted on average by participants in this experiment. In addition, standard error of this mean is calculated as 0.197 which shows that statistically this number can be used as a threshold \cite{durao2012personalized}. 
	
\begin{figure}[h!]
  \includegraphics[width=\figwidth]{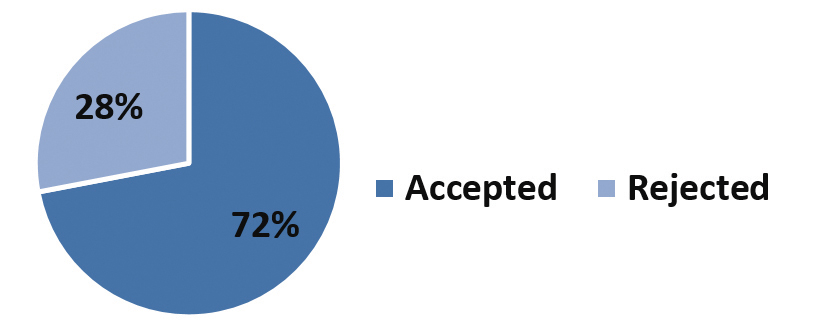}
  \caption{Percentage of accepted and rejected websites in total (5 Recommendations)}
  \label{result-5-2}
\end{figure}

	When all recommendations are considered, it can be seen from Figure~\ref{result-5-2} that 72 \% of the recommended websites are accepted in total.

\begin{figure}[h!]
  \includegraphics[width=\figwidth]{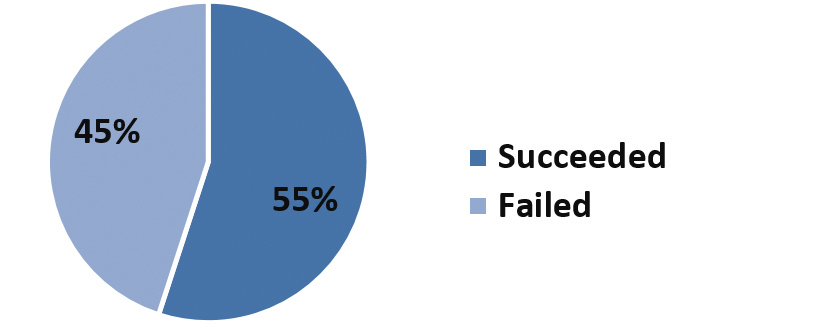}
  \caption{Percentage of succeeded and failed users (5 Recommendations)}
  \label{result-5-3}
\end{figure}

    In addition in Figure~\ref{result-5-3}, percentage of users that can be counted as succeeded according to the threshold of 3.6 can be seen as 55 \%. This shows that more than half of the users have received acceptable level of recommended websites.  
    
\begin{figure*}
  \includegraphics[width=\textwidth]{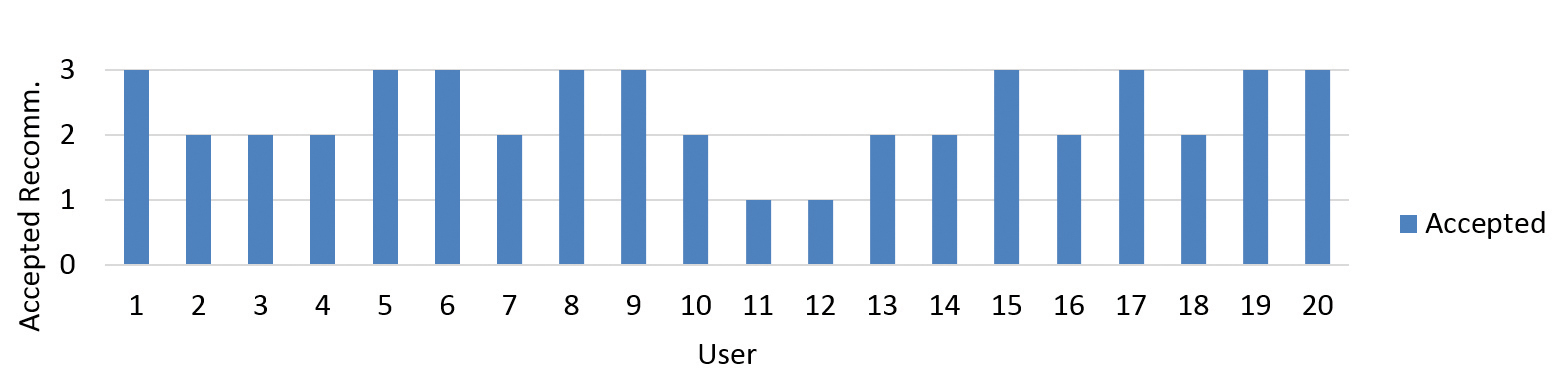}
  \caption{Accepted recommendations by each user (3 Recommendations)}
  \label{result-3-1}
\end{figure*}

As mentioned in Section 4.2, in this experiment level of acceptances of top 3 websites are also tracked because in data gathering stage only five websites were asked from participants. Considering this low number of input, high success on the same number of recommended outputs cannot be expected. Therefore, performance of this model on a smaller output set of 3 websites is also analyzed. When recommended websites for each user are analyzed, it can be seen from the Figure~\ref{result-3-1} that at least 1 of 3 recommendations is accepted. In addition, 9 of the 20 users have accepted all of 3 recommended websites.

\begin{figure}[h!]
  \includegraphics[width=\figwidth]{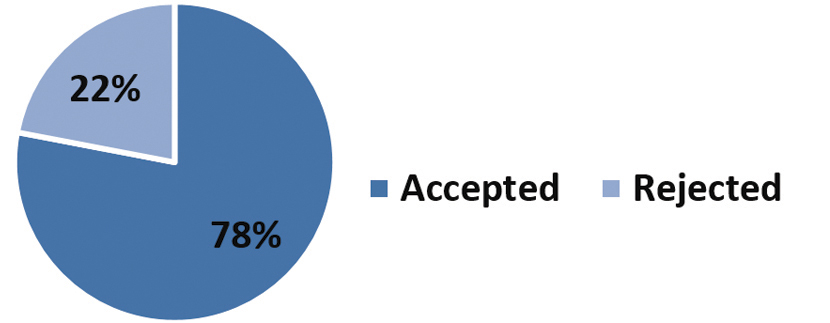}
  \caption{Percentage of accepted and rejected websites in total (3 Recommendations)}
  \label{result-3-2}
\end{figure}

    When acceptance level is further analyzed, on the average 2.35 of 3 websites are accepted with a standard error of mean of 0.15. Although this low error shows that this mean can be used as a threshold, it will not provide more and beneficial information because only the all-accepted users will be able to pass this threshold. On the other hand, percentage of accepted websites increased to 78 \% in this case as shown in Figure~\ref{result-3-2}. This increase is a natural expectation because in this case instead of 5 most similar websites, top 3 of them are sent.

When the overall performance of this model is taken into account, it yielded better results than expected results. In other words, this model was not able to reach excellent acceptance level but achieved way more than satisfactory. However, before implementing this method on real life, more comprehensive experiments must be undertaken.

\subsection{Discussion}

When the presented results are evaluated, it can be said that the method performs well in recommending new websites or catching users’ current interests. However, when the recommended websites are analyzed some important points are worth mentioning. Firstly, there is no control on the tags provided by users in this system. Although users do not intend to mislead the method while tagging websites, different purposes of tagging can create such a problem. In addition, when some of these different purpose tags happen to be one of the popular tags; similarity of non-similar websites stands out. Secondly, meanings of the tags are added as they are user inputs in this system. This created an environment where each user tagged the documents with all possible meanings that may be used by other users. However, it is open to debate whether or not the meanings should be as important as original tags. Because in some cases synonyms of tags can relate non similar documents and result with unaccepted recommendations. In order to solve this problem, different weights can be used for synonyms of the tags before including into data. When these drawbacks of the model are considered with the different expectations of users, nearly 30 \% of rejected recommendations can be explained.

\section{Conclusion}

In this paper a Turkish-language tag-based recommendation system which is based on similarity, tag weight, tag popularity; where semantic properties of tags are taken into account is presented. Contribution of this paper was combining well-known similarity measures and calculations with a Turkish semantics analysis. In order to evaluate the model, an experiment with 25 people is undertaken where participants are supposed to provide websites and tags; and then evaluate recommendations.

\section{Future Work}

As further studies related to the method presented in this paper, two possible improvements are proposed for different stages.

Firstly, in preprocessing stage spell checking and stemming operations are undertaken. However, no other control is made on tags. When user inputs are analyzed, although Turkish inputs are asked, widely used English words are seen in dataset. For instance, "e-mail" is provided as a tag which is a common usage in Turkish but not actually a word in Turkish. In addition, due to different keyboard layouts, some users can provide Turkish tags in English letters, for instance with "i" instead of "{\i}". Therefore, some kind of control or translation can be implemented in preprocessing stage.

Secondly, semantic analysis of this method uses relatively small set of synonyms and synonyms of the tags are counted as important as original inputs. Therefore, a more comprehensive thesaurus for Turkish can be implemented with weights so that synonyms of the tags will be less important than the original tags.



\section{References}
\label{}

 
\bibliographystyle{elsarticle-num}

\bibliography{references}
 
\end{document}